# Computational models of astrocyte function at glutamatergic synapses

**Authors:**
Kerstin Lenk [1,2,3,*]
Audrey Denizot [4,5,6,*]
Barbara Genocchi [7]
Ippa Seppälä [8]
Marsa Taheri [9]
Suhita Nadkarni [10]

**Affiliations:**
[1] Institute of Neural Engineering, Graz University of Technology, Graz, Austria
[2] BioTechMed, 8010 Graz, Austria
[3] Theoretical Sciences Visiting Program, Okinawa Institute of Science and Technology Graduate University, Onna, Japan
[4] Okinawa Institute of Science and Technology, Computational Neuroscience Unit, Onna-Son, Japan
[5] AIstroSight, Inria, Hospices Civils de Lyon, Université Claude Bernard Lyon 1, Villeurbanne, France
[6] University of Lyon, LIRIS UMR5205, Villeurbanne, France
[7] Faculty of Medicine and Health Technology, Tampere University, Tampere, Finland
[8] Department of Psychology and Speech-Language Pathology, University of Turku, Turku, Finland
[9] Department of Neurobiology, University of California Los Angeles, CA, USA
[10] Department of Biology, Indian Institute of Science Education and Research Pune
* Kerstin Lenk and Audrey Denizot contributed equally to this work.

**Email address of the corresponding author:** lenk.kerstin@gmail.com

**Summary/Abstract**

At tripartite synapses, astrocytes are in close contact with neurons and contribute to various functions, from synaptic transmission, maintenance of ion homeostasis and glutamate uptake to metabolism. However, disentangling the precise contribution of astrocytes to those phenomena and the underlying biochemical mechanisms is remarkably challenging. This notably results from their highly ramified morphology, the nanoscopic size of the majority of astrocyte processes, and the poorly understood information encoded by their spatiotemporally diverse calcium signals. This book chapter presents selected computational models of the involvement of astrocytes in glutamatergic transmission. The goal of this chapter is to present representative models of astrocyte function in conjunction with the biological questions they can investigate.

# 1. Introduction

In this chapter, we present diverse computational models of astrocytes on a wide range of spatiotemporal scales. Our goal is to equip the reader with a concise overview of the available models so that they can use them to investigate research questions of interest. Astrocytes are glial cells that are essential to numerous functions of the central nervous system, from brain development, metabolism, and homeostasis to brain injury repair. They interact with numerous cell types simultaneously. Notably, astrocytes communicate with both blood vessels at specialized subcellular compartments – endfeet – and with neurons at tripartite synapses, where the astrocyte is in apposition to presynaptic and postsynaptic elements *(1)*. This intercellular communication is believed to be mediated by astrocyte $Ca^{2+}$ signals *(2)*. One of the largest challenges in the field is to unravel the biochemical reactions that underlie astrocyte function *(3).* Because of the complex nanoscopic morphology of astrocytes and the various spatiotemporal properties of astrocyte signals, characterizing their involvement in the brain (patho-)physiology is hindered by technical difficulties, such as the resolution of live microscopy *(2, 4)*. Computational models of astrocytes have been implemented to overcome those limitations and have provided valuable insights into the involvement of astrocytes in synaptic function *(5–7)*.

A computational model is a simplification of a system that describes its elements, their states, and their interactions. Computational models can guide experimentalists towards the most relevant experiments by forming a theoretical framework to characterize and predict the function of the system of interest *(5, 8, 9)*. Those models can be used to run studies that would be unfeasible or time-consuming experimentally, in a fully controllable manner. Computational models thus make it possible to go beyond correlational observations and to propose causal relationships that govern the dynamics of the system of interest. When data is rare or difficult to get experimentally, models can be used to generate vast amounts of synthetic realistic datasets.

Biophysical models *(10)* describe detailed chemical reactions based on experimental data of a biological phenomenon. The advantage of this type of modeling paradigm is that it can provide quantitative, testable predictions. The potential disadvantage of this approach can be its high computational cost, i.e. a large amount of time or memory required to run the simulations. Another type of approach is phenomenological modeling (the FitzHugh-Nagumo model *(11)* is a good example), which aims at mimicking the phenomenon of interest without describing its biophysical details. Those models are less computationally-expensive, allowing simulations of larger systems and/or reduced simulation time. For more details on the insights gained by experimentalists from computational models and vice versa, the reader can refer to *(12)* and to dedicated books *(13, 14)*.

Several reviews of computational astrocyte models have been published recently that have contributed to a better understanding of astrocyte function in the brain *(5–7)*, and new ones emerge at a fast pace. The review paper by Oschmann et al. *(5)* summarizes astrocyte models from the subcellular to the network level. The book chapter from Denizot et al. *(6)* presents the recent developments in astrocyte modeling approaches at the cellular and subcellular levels that accompanied advances in experimental techniques. Manninen et al. *(7)* provide a historical overview of computational astrocyte models. Lastly, the book "Computational Glioscience" *(15)* gives detailed insights into various computational models of neuron-glia interactions. We believe that robust and insightful collaborative work between theoreticians and experimentalists relies on the implementation of models that are accessible, reusable (FAIR principles: Findable, Accessible, Interoperable, Reusable) *(16)*, as well as reproducible and replicable *(8, 17, 18)*. In this line, we provide links to the code of each publicly available model (Table 1) and describe the validation of the model against experimental data whenever the latter was performed.

In this chapter, rather than presenting an exhaustive list of astrocyte models, we present a few selected models that are most representative of the diversity of the existing models and highlight the type of biological questions they can investigate. The chapter is organized as follows: we present models of glutamatergic transmission at the tripartite synapse (Section 2), the involvement of astrocytes in glutamate uptake (Section 3), ion homeostasis (Section 4), and metabolism (Section 5). Then, we introduce models that study astrocyte structure-function coupling (Section 6) and astrocyte networks (Section 7). We end the chapter with concluding remarks (Section 8) and a list of resources (Section 9).

# 2. Signal transmission at tripartite synapses

Astrocytes can contact pre- and postsynaptic neurons, forming so-called tripartite synapses *(1)*. The astrocyte subcompartments that communicate with neurons at synapses are often referred to as perisynaptic astrocyte processes (PAPs). Glutamate released by active glutamatergic presynaptic neurons binds to G-protein-coupled receptors at the astrocyte membrane, which triggers a series of chemical reactions that allow the formation of $Ca^{2+}$ signals in the astrocyte cytosol. Those $Ca^{2+}$ signals are essential to various brain functions, from synaptogenesis to memory consolidation *(19, 20)*. They are believed to have a variety of downstream effects that can modulate neurotransmission and synaptic plasticity, such as the release of gliotransmitters (e.g., adenosine triphosphate (ATP), D-Serine, and glutamate) by the astrocyte into the synaptic cleft *(21)* or changes in extracellular ion concentrations (more on this in Section 4). Gliotransmission has long been debated *(22, 23)* and a growing body of literature supports the existence of such neuron-astrocyte communication *(24, 25)*. The influence of astrocyte $Ca^{2+}$ signals and astrocyte-neuron communication on synaptic transmission and plasticity are still poorly understood. Computational models have been developed to study those interactions and to predict the coupling between neuronal and astrocyte activity.

In this section, we present three models of signal transmission at tripartite synapses *(26–28)*, highlighting the different approaches and strategies that have been used as well as the different insights that can be gained from those models (Figure 1). As glutamate uptake and ionic homeostasis are covered in Sections 3 and 4, we here focus on models of neuronal activity-induced astrocyte $Ca^{2+}$ dynamics and gliotransmission. Those models can be used to predict how astrocyte activity can modulate neurotransmission in various physiological conditions.

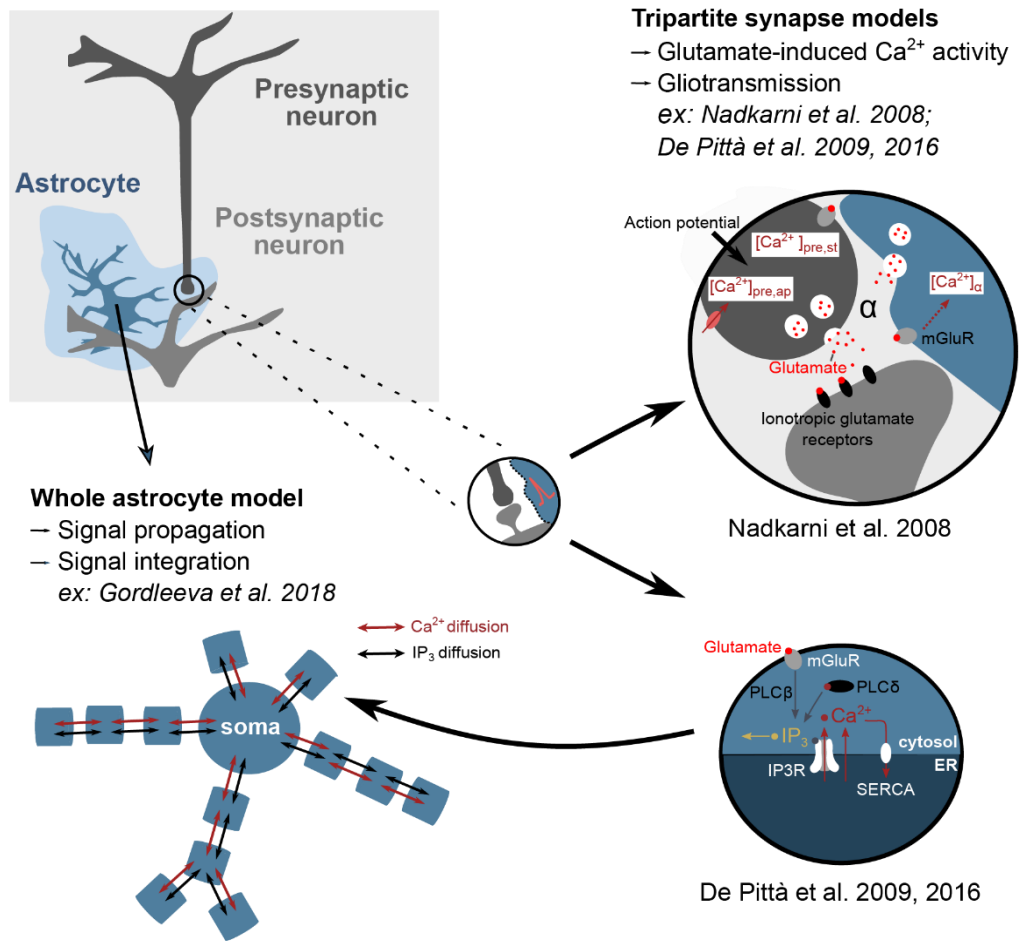

**Figure 1: Models of signal transmission at tripartite synapses.** Neuron-astrocyte communication at synapses has been modeled using different strategies: simulating neuronal activity-induced $Ca^{2+}$ activity in astrocytes with (e.g., Gordleeva et al. model *(26)*) or without (e.g., De Pittà et al. model *(27)* and Nadkarni et al. model *(28)*) taking into account the diffusion of molecules between subcellular compartments.

## 2.1. Model of glutamate-induced IP$_3$ and Ca$^{2+}$ oscillations in astrocytes

In 2009, De Pittà et al. *(27)* introduced a model in which the astrocyte receives neuronal glutamate input. Glutamate then binds to the metabotropic glutamate receptors (mGluRs) at the astrocyte membrane. This binding results in an increase in inositol trisphosphate (IP$_3$) concentration in the astrocyte cytosol. Subsequently, $Ca^{2+}$ is released from the endoplasmic reticulum (ER) through the opening of IP$_3$ receptor channels (IP$_3$Rs), which leads to IP$_3$-dependent $Ca^{2+}$-induced $Ca^{2+}$ release (CICR) in the cytosol, whereby an increased $Ca^{2+}$ concentration enhances the release of $Ca^{2+}$ from the ER. At high $Ca^{2+}$ concentrations, the IP$_3$Rs are inactivated and $Ca^{2+}$ is pumped back into the ER by the sarco-endoplasmic reticulum $Ca^{2+}$-ATPase (SERCA) pumps. A long-lasting glutamate stimulus leads to high cytosolic $IP_3$ concentrations, allowing for the alternation of activation and inactivation of $Ca^{2+}$ channels, and thus $Ca^{2+}$ oscillations *(29)*. The kinetics of IP3Rs are described with the Li-Rinzel model *(30)* (reviews of $IP_3R$ models can be found in *(31)* and *(32)*). Besides the CICR mechanism, $IP_3$ production by the phospholipase C (PLC) isoenzymes PLCβ and PLCδ and $IP_3$ degradation are described by the model. $IP_3$ degradation can occur in two ways, through dephosphorylation by inositol polyphosphate 5-phosphatase and $Ca^{2+}$-dependent phosphorylation by $IP_3$ 3-kinase. A simplified representation of the modeled chemical reactions can be found in Figure 1. In this paper, the authors measured the variation of the amplitude and frequency of $Ca^{2+}$ signals depending on the level of synaptic activity, which is modeled as alterations of the glutamate concentration in the extracellular space (ECS).

In summary, the model describes $IP_3R$-dependent $Ca^{2+}$ signals in the astrocyte that result from neurotransmission at glutamatergic synapses. It has been used in various other models to describe astrocyte activity, e.g., *(26, 33–35)*.

## 2.2. Model of $Ca^{2+}$ activity in an astrocyte connected to a neuronal network

To model astrocyte-neuron communication at tripartite synapses, Gordleeva et al. *(26)* simulated the activity of a single astrocyte in a network of 36 neurons. Their goal was to investigate the effect of the spatial distribution of $Ca^{2+}$ signals within the astrocyte on gliotransmitter release and the associated modulation of neuronal activity. To do so, the authors combined their compartmental model of an astrocyte *(36)* with their model of a neuronal network, which accounts for gliotransmitter release by astrocytes *(37)*. Astrocyte morphology is described as the assembly of cellular subcompartments, with a cylindrical shape, coupled by $IP_3$ and $Ca^{2+}$ diffusion (deterministic spatially-extended approach, see Section 6 for more details). The astrocyte contains 14 processes, each connected to a different synapse amongst the 36 neurons of the network. Astrocyte $Ca^{2+}$ activity in each process is modeled using the $IP_3R$-mediated $Ca^{2+}$ signaling model from De Pittà et al. *(27)* (Section 2.1). In the model from Gordleeva and colleagues, gliotransmitter release is a function of cytosolic $Ca^{2+}$ concentration in distal processes and only occurs when local $Ca^{2+}$ concentration exceeds a given threshold. The neurons are modeled using a conductance-based mathematical formalism, the Hodgkin-Huxley model *(38)*. The connectivity of the neural network is randomly chosen, with a 20 % probability of connectivity for each pair of neurons. Each spike in a presynaptic neuron results in the release of glutamate at synapses, modeled by a Poisson process of fixed frequency. This model allows one to simultaneously monitor $Ca^{2+}$ signals in different astrocyte compartments, gliotransmitter release at synapses, and postsynaptic neuronal firing. The modeled geometry and reactions are summarized in Figure 1.

To investigate the connectivity between active neurons and astrocytes, Gordleeva et al. *(26)* have calculated the cross-correlation between neuronal firing rate and the so-called astrocyte firing rate, corresponding to the frequency of the number of $Ca^{2+}$ signals in the whole astrocyte. They found a synchronization of the activity of presynaptic neurons and astrocytes with a delay of roughly 2 seconds. They further analyzed the integration of $Ca^{2+}$ signals within the astrocyte and predicted that distal processes were the most active subcompartments of astrocytes, occasionally allowing somatic events to occur, the latter backpropagating to all processes. The model was able to predict various effects of astrocyte gliotransmitter release on neuronal activity: glutamate-mediated presynaptic potentiation, inhibition of presynaptic release, and D-Serine-mediated increase in the postsynaptic current. The authors also investigated the effect of local $Ca^{2+}$ signals in a PAP contacting an active synapse on neighboring processes. They found that $Ca^{2+}$ and $IP_3$ diffusion within the astrocyte can activate the release of gliotransmitters in neighboring processes, resulting in the potentiation or depression of nearby inactive synapses. Similarly, whole-cell $Ca^{2+}$ events favored the release of gliotransmitters from various processes and thus coordinated the activity of the neural circuits connected to the active astrocyte.

Overall, the model from Gordleeva et al. *(26)* accounts for the spatial morphology of a single astrocyte and its connectivity to various neurons. It is thus well-suited to study the modulation of neuronal network activity mediated by the spatiotemporal integration of $Ca^{2+}$ signals within a single astrocyte. Please refer to Section 7 for network models involving numerous astrocytes.

## 2.3. Model of signal transmission at a glutamatergic tripartite synapse

Nadkarni et al. *(28)* modeled the closed-loop modulation of synaptic transmission at a tripartite synapse in the hippocampus (Figure 1). The model is based on the Bertram et al. phenomenological model to describe action potential-driven vesicle release from the presynaptic neuron *(39)*. Both asynchronous and action potential-independent neurotransmitter release rates are derived from experimental data *(40)*. This release is followed by a refractory period *(41)*. The vesicle recycling rates and short-term plasticity are determined by the Tsodyks et al. model *(42)*. Thus, the Nadkarni et al. model considers both the availability and recovery of neurotransmitter resources. Furthermore, the model describes the $Ca^{2+}$ activity of an astrocyte, the resulting release of gliotransmitters (here glutamate), and its effect on synaptic transmission. If a vesicle release event takes place, the glutamate in the cleft binds to the postsynaptic receptors and the mGluRs on the membrane of the astrocyte process. The former causes an excitatory postsynaptic current, while the latter results in the production of $IP_3$, causing the release of $Ca^{2+}$ from the astrocyte ER (modeled as per Nadkarni et al. *(43)*). The time course of $Ca^{2+}$ signals in the astrocyte in the model is in qualitative agreement with experimental data from rat hippocampal and visual cortex slices *(44)*. If the cytosolic $Ca^{2+}$ levels in the astrocyte are above a threshold, glutamate is released from the astrocyte *(45, 46)*, which leads to the potentiation of synaptic transmission *(40, 47, 48)*. Since the precise mechanisms of gliotransmitter release by the astrocyte were unknown, the authors modeled gliotransmission with a slow decay that mirrors the timescale of potentiation mediated by astrocytes *(48)*.

In their study, Nadkarni et al. propose that $Ca^{2+}$ release from the astrocyte leads to a potentiation of neurotransmitter release that can last for minutes. In support of the proposed mechanism, the time course of the ER-mediated $Ca^{2+}$ signal correlates well with the observed time course of changes in synaptic transmission. Since these $Ca^{2+}$ fluxes are not temporarily correlated to the action potential-mediated $Ca^{2+}$ activity, it causes an increase in asynchronous release, i.e. in action potential-independent neurotransmitter release, which depletes the vesicle resource. The strength of the coupling between the astrocyte $Ca^{2+}$ and the presynaptic $Ca^{2+}$ was investigated as an open parameter, '**α**'. An increase in the value of '**α**' thus leads to an enhanced vesicle release rate. The authors found a value of '**α**' that resulted in a neurotransmitter release rate that was concordant with experimental data *(47)*. Interestingly, it is the value of '**α**' and the corresponding increase in release probability that seemed to maximize synaptic transmission. The model was robust with respect to a wide range of stimulus frequencies, the number of active zones, and basal levels of vesicular release probability.

In summary, the model of Nadkarni et al. describes neuron-astrocyte coupling at tripartite synapses and can be used to investigate the complex relationship between astrocyte activity, presynaptic vesicular release rate, and vesicle depletion. It can be used to predict the modulation of neurotransmitter release rate by astrocytes under a range of stimulus protocols and can be extended to calculate downstream changes in plasticity.

## 2.4. Discussion

Numerous models have been implemented to study the roles of astrocytes in synaptic transmission. In those models, astrocytes mainly respond to neuronal activity through the mGluR-dependent activation of $IP_3R$ channels that influence $Ca^{2+}$ dynamics. De Pittà et al. *(27)* modeled astrocyte $IP_3$ and $Ca^{2+}$ oscillations mediated by glutamatergic transmission; Nadkarni et al. *(28)* described the coupling of neuronal and astrocyte activity in a single tripartite synapse; while Gordleeva et al. *(26)* depicted the spatiotemporal integration of astrocyte $Ca^{2+}$ signals in a whole cell, thus predicting its influence on synaptic transmission at various synapses. All three models (Figure 1) describe $IP_3R$ channel dynamics using early computational models *(30, 49)* that are based on electrophysiological data such as Bezprozvanny et al. *(50)*.

# 3. Glutamate uptake

As much as glutamate release is essential for excitatory transmission in the Central Nervous System (CNS), its rapid removal from the ECS is critical for normal brain function. Glutamate molecules that linger on and diffuse away from the synaptic cleft can compromise the specificity of synaptic signaling *(51)*, a key component of information processing in the brain. A prolonged lifespan of glutamate can also cause neuronal cell death through a phenomenon referred to as excitotoxicity *(52)*.

About 90% of all the released glutamate is taken up by astrocytes, making them the primary cells responsible for glutamate clearance *(53–55)*. This phenomenon is mediated by its uptake by glutamate transporters, which are expressed in all cell types in the CNS, with the highest density found in astrocytes *(56)*. Glutamate transporters can be classified as $Na^+$-independent and $Na^+$-dependent transporters *(56, 57)*. Even though the affinity of $Na^+$-independent transporters is similar to that of $Na^+$-dependent ones, they contribute to less than 5% of the total glutamate uptake *(56)*. $Na^+$-dependent glutamate transporters are also referred to as excitatory amino acid transporters (EAATs) and consist of several isoforms, e.g., EAAT-1 (also known as glutamate-aspartate transporter or GLAST in rodents) and EAAT-2 (known as glutamate transporter-1 or GLT-1 in rodents). The time course of glutamate uptake can be calculated based on a deconvolution analysis of astrocyte transporter currents such as done by Scimemi and Diamond *(58)*. In this section, we present three models of astrocyte glutamate uptake *(59–61)* to study its effect on astrocyte $Na^+$ and $Ca^{2+}$ dynamics and postsynaptic α-amino-3-hydroxy-5-methyl-4-isoxazole propionic acid receptors (AMPAR) and NMDAR activation (Figure 2).

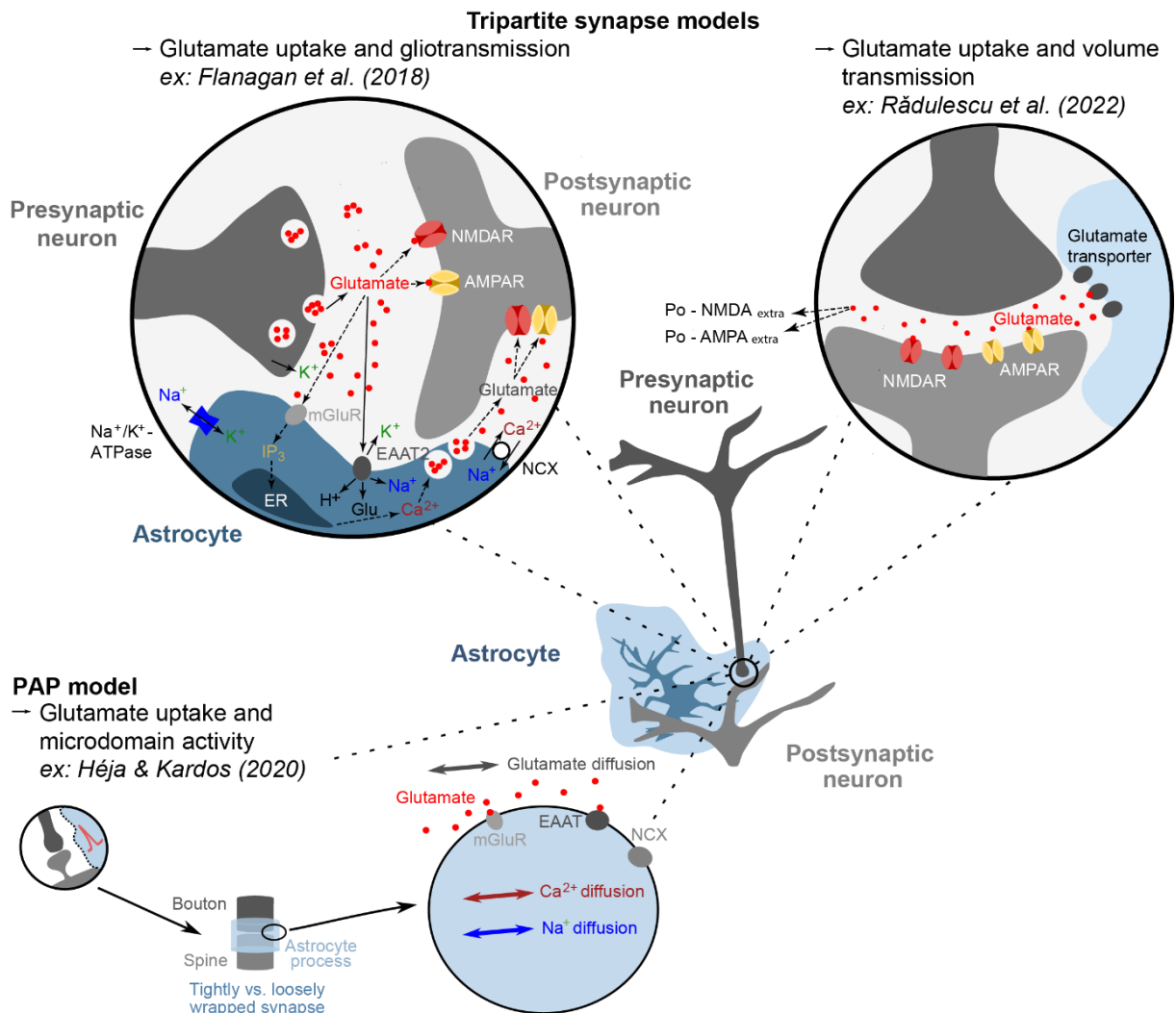

**Figure 2: Models of glutamate uptake at tripartite synapses.** Flanagan et al. *(60)* and Rădulescu et al. *(59)* model the interplay between glutamate uptake by astrocytes at tripartite synapses and postsynaptic activity. Additionally, Flanagan et al. describe how those interactions can affect gliotransmission. Rădulescu et al. instead simulate the influence of glutamate uptake on extrasynaptic volume transmission. The Héja & Kardos model *(61)* describes how glutamate uptake alters astrocyte microdomain activity at the single PAP level.

## 3.1. Model of glutamate uptake in relation to glutamate transporter density

A recent model proposed by Rădulescu et al. *(59)* investigates the effect of the density of glutamate transporters on the membrane of PAPs on the opening probability of AMPARs and NMDARs at the active (Po-$AMPA_{peri}$ and Po-$NMDA_{peri}$) and neighboring synapses (Po-$AMPA_{extra}$ and Po-$NMDA_{extra}$). To do so, they performed Monte Carlo simulations (particle-based approach, see Section 6 for more details), in which glutamate molecules were injected at the center of a synaptic cleft and the glutamate concentration profile was monitored at the stimulated synapse as well as at six neighboring synapses (Figure 2). Glutamate transporter density in astrocytes at different ages in mice (2 weeks to 21 months) was estimated based on dot blot experiments. Combining those values with 3D axial STEM tomography reconstructions allowed the authors to evaluate the variability of glutamate transporter density in different astrocyte subcellular compartments.

Simulations with the model allowed the authors to infer the effect of the variability of glutamate transporter density on synaptic transmission depending on mouse age and the subcellular compartment in contact with the synapse. Results indicate that the density of glutamate transporters on PAPs as well as the location of the site of glutamate release play an important role in shaping glutamate receptor activity at local (Po-AMPA$_{peri}$ and Po-NMDA$_{peri}$) and distant (Po-AMPA$_{extra}$ and Po-NMDA$_{extra}$) synapses. More specifically, the higher the glutamate transporter density around a synapse, the lower the extra-synaptic activation of glutamate transporters.

In summary, this model allowed the authors to test the effect of the differential expression levels of glutamate transporters that they observed in different astrocyte compartments on the local and extra-synaptic activation of NMDARs and AMPARs. Their study also highlights the importance of accurate estimates of molecular expression levels to fully grasp the spatiotemporal dynamics of glutamatergic signaling. This model can be used to gain insights into glutamate concentration profiles in the ECS at tripartite synapses, glutamate receptor activity, and glutamate spillover under a range of stimulation protocols.

## 3.2. Model of glutamate uptake by astrocytes and its effects on post-synaptic neuronal excitability

Flanagan et al. *(60)* extended the tripartite synapse model by De Pittà and Brunel *(62)* by adding a biophysical model of EAAT-2 transporter activity at the astrocyte membrane (Figure 2). Thus, they combined in their model glutamate uptake and gliotransmission. Additionally, the model also accounts for Na$^+$ (as EAAT-2 is a Na$^+$-dependent glutamate uptake transporter) and K$^+$ dynamics. The model includes five compartments: the presynaptic and postsynaptic terminals, the soma and process of an astrocyte, and the ECS. The phenomenological presynaptic terminal releases glutamate at a given rate and spikes trigger a K$^+$ efflux from the presynaptic neuron into the synaptic cleft. The postsynaptic terminal is populated with AMPARs and NMDARs. The binding of glutamate to mGluRs on the membrane of astrocytes results in the activation of IP$_3$Rs on the ER and leads to Ca$^{2+}$ release into the cytosol (see Section 2). This Ca$^{2+}$ signal can cause a release of gliotransmitter from the perisynaptic compartment of the astrocyte. Moreover, glutamate is taken up by EAAT-2, which is accompanied by a Na+ influx into the astrocyte cytosol. The Na$^+$/K$^+$-ATPase pump is primarily responsible for maintaining Na$^+$ and K$^+$ gradients across the cell membrane; while the Na$^+$-Ca$^{2+}$ exchanger (NCX) removes Ca$^{2+}$ from the cell and allows Na$^+$ influx into the cell.

Flanagan et al. simulated three basal glutamate concentrations in the astrocyte. The highest of those concentrations (10 mM), non-physiological and chosen to simulate pathological conditions, led to a delayed removal of glutamate by EAAT-2 from the synaptic cleft. They hypothesize that this slow removal also occurs in the epileptic brain. Due to the prolonged glutamate uptake, the activation of mGluRs was higher, which led to larger $IP_3$-mediated $Ca^{2+}$ oscillations, allowing intracellular $Ca^{2+}$ concentration to rise above the threshold for gliotransmission. This increased glutamate release by the astrocyte triggered a slow inward current, which resulted in high-frequency neuronal activity. Interestingly, such a high intracellular glutamate concentration in the astrocyte reduced the minimum value of the neuronal firing rate that could trigger gliotransmitter release events.

In summary, the model allows inferring the influence of glutamate uptake and astrocyte intracellular glutamate concentration on gliotransmission and the subsequent alterations of postsynaptic firing rates. It is well-suited to study the interplay between astrocyte activity and the excessive glutamate concentrations measured in the ECS in pathological conditions, such as epilepsy.

## 3.3. A spatial model of glutamate uptake, $Ca^{2+}$ and $Na^+$ signaling in PAP microdomains

Recently, Héja & Kardos *(61)* have developed a model at the nanoscale level to investigate how glutamate uptake as well as $Na^+$ and $Ca^{2+}$ activity in the astrocyte cytosol are affected by the coverage of the synapse by the astrocyte leaflet. Due to the small size of PAPs, the model is both stochastic and spatially extended (see Section 6 for more details). Briefly, the model describes glutamate diffusion within the synaptic cleft as well as the diffusion of $Na^+$ and $Ca^{2+}$ within the astrocyte cytosol. The geometry of the synapse is simplified: the presynaptic bouton and the postsynaptic spine are cylinder-shaped while the astrocyte process consists of a hollow cylinder. Astrocyte process geometries of different sizes are used, enabling to test the effect of the astrocyte coverage of the synapse – from loose to tight – on glutamate uptake (see Figure 2). Neuronal activity is modeled as a punctual infusion of 5,000 glutamate molecules at the center of the presynaptic bouton. The model describes the glutamate uptake by EAATs at the plasma membrane of the astrocyte process, which is accompanied by an influx of $Na^+$ within the astrocyte cytosol. EAATs can interact with any glutamate molecule in their vicinity, here described as a 50x50x50 $nm^3$ interaction space. The model also takes into account the activity of the NCX, which is the only source of $Ca^{2+}$ influx and efflux in this model.

The model by Héja & Kardos has contributed to a better understanding of the dynamics of glutamate uptake at individual synapses. They distinguished a sub-population of EAATs on the membrane of astrocytes in the tightly-wrapped tripartite synapse configuration that was exposed to high glutamate concentrations and was responsible for the majority of glutamate uptake. Interestingly, this sub-population was absent in the loosely-wrapped synapse configuration. They were further able to characterize the fluctuations of $Ca^{2+}$ concentration resulting from $Ca^{2+}$ binding and unbinding to and from NCX and showed that those fluctuations were remarkably stable, displaying little variability upon changes of $Na^+$ and $Ca^{2+}$ concentrations, neuronal firing rate, NCX activity, or membrane potential.

Overall, the model from Héja and Kardos is an example of a model focusing on astrocyte-neuron communication at the single astrocyte process level. Such a model is well-suited to investigate the effect of spatial properties, such as the shape of the astrocyte leaflet (see also Section 6), on astrocyte $Ca^{2+}$ microdomain activity, and glutamate uptake.

## 3.4. Discussion

Over the last few decades, experimental and computational studies have characterized the biophysical properties and expression levels of astrocyte glutamate transporters in various (patho-)physiological conditions, in the mature and developing central nervous system. The computational models presented in this section can be used to further our understanding of how astrocyte transporters corral glutamatergic transmission (Flanagan et al. *(60)* and Héja & Kardos *(61)* models) and limit glutamate spillover (Rădulescu et al. model *(59)*).

# 4. Ion homeostasis

Astrocytes express numerous ion transporters, pumps, exchangers, channels, and receptors that regulate ion homeostasis and the ECS volume *(3)*. Here, we describe three computational models of astrocyte ion homeostasis that provide key insights into the complex interactions between ion fluxes, ECS shrinkage, glutamate uptake, gliotransmission, and/or astrocyte $Ca^{2+}$ signaling (Figure 3). The most common ion pumps and transporters described in these models *(63–65)* are i) $Na^+/K^+$-ATPase pumps, which actively exchange $Na^+$ (outwards) and $K^+$ (inwards) across the plasma membrane; ii) NCX exchangers, which exchange $Na^+$ (outwards) and $Ca^{2+}$ (inwards) – NCX can switch to reverse mode with high intracellular $Na^+$ concentrations; iii) EAAT-1/2, which exchange $Na^+$ and glutamate (both inwards) with $K^+$ (outwards). Each astrocyte subcompartment, from the soma to PAPs and endfeet, display different levels of expression of transporters, pumps, exchangers, channels, and receptors *(3, 59)*. Hence, the molecular processes underlying ion homeostasis might vary across the astrocyte. This is particularly difficult to examine experimentally as astrocyte fine processes, which account for about 75% of the astrocyte volume *(66)*, are not resolved by diffraction-limited light microscopy.

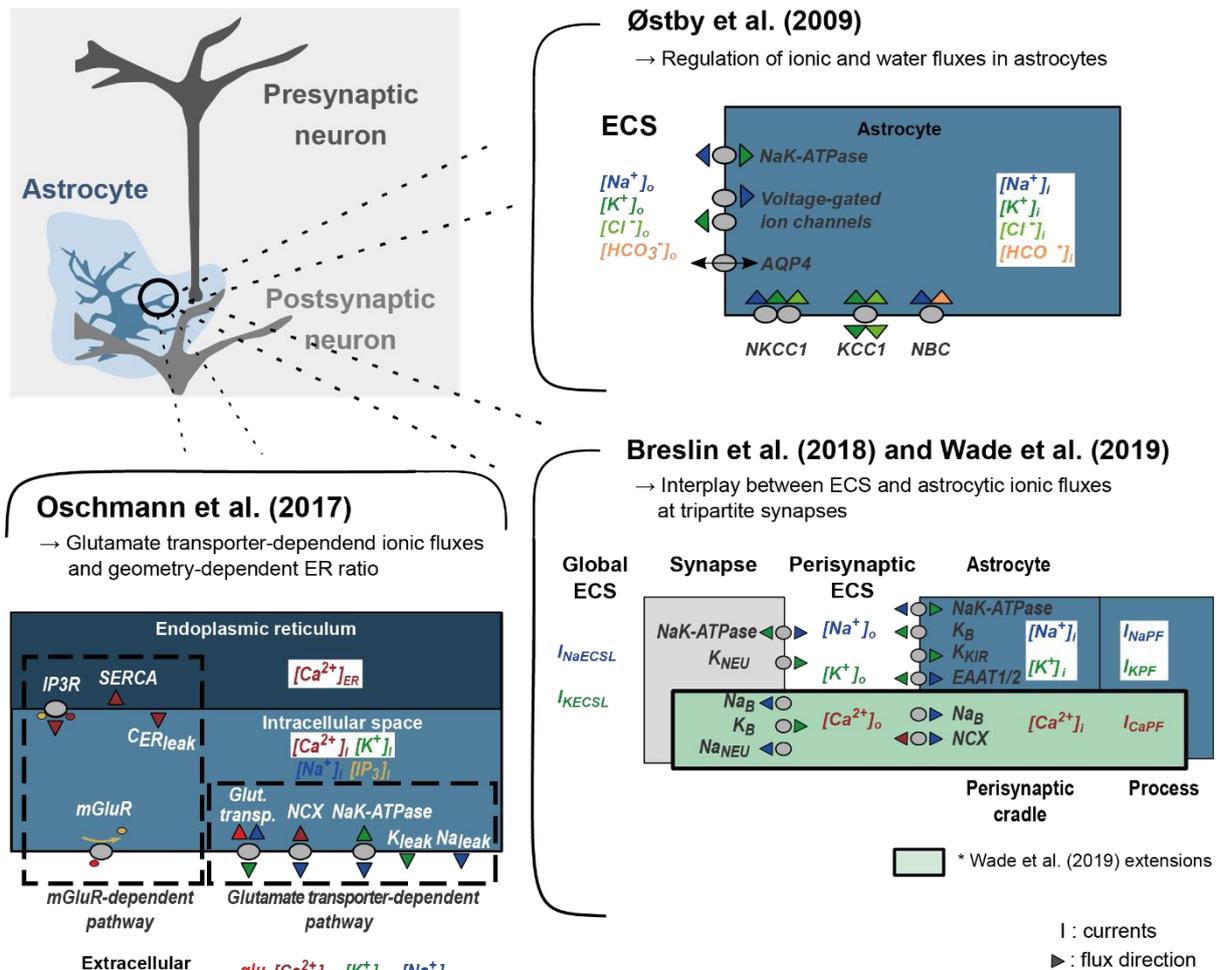

**Figure 3: Models of ion homeostasis at tripartite synapses.** Østby et al. *(63)* described the ECS shrinkage resulting from astrocyte ionic fluxes. The Breslin et al. model *(64)* comprises a synapse (describing both pre- and postsynaptic neurons) encapsulated by an astrocyte process. Wade et al. *(67)* extended the model by adding $Ca^{2+}$ and $Na^+$ dynamics. The Oschmann et al. model *(65)* describes the interplay between mGluR-dependent and glutamate transporter-dependent $Ca^{2+}$ signaling pathways in the astrocyte. The colored arrowheads describe the direction of $Na^+$ (blue), $K^+$ (light red), $Ca^{2+}$ (dark red), $Cl^-$ (green), and $HCO_3^-$ (orange) fluxes.

### 4.1. Model of astrocyte ion fluxes-mediated ECS shrinkage

Østby et al. *(63)* developed a computational model to study the interplay between $K^+$ buffering and water transport mechanisms. Ionic transport across the membrane change the relative amount of positive and negative charges in the intracellular space (ICS) and the ECS. These fluxes of charges modulate the intracellular osmolarity that drives water fluxes at the plasma membrane. The model accounts for the variations in $Na^+$, $K^+$, and bicarbonate ion ($HCO_3^-$) concentrations in the astrocyte and the ECS. Those ion concentration changes are described, together with the volumetric changes of the astrocyte and the ECS, using ordinary differential equations (ODEs), while algebraic equations depict $Cl^-$ dynamics and astrocyte membrane potential. The authors implemented the model to investigate stimulation-induced shrinkage of the ECS in the gray and white matter. When active, neurons release $K^+$ and uptake $Na^+$, while astrocytes uptake $K^+$, $Na^+$, and $Cl^-$, which results in a water influx from the ECS into the cell by osmolarity. Figure 3 displays the modeled reactions.

The simulation results indicated that volume changes are controlled by the combined action of several processes. The ECS shrinkage seemed to be enhanced by the cotransporters (i.e., the $Na^+$-bicarbonate cotransporter, NBC, and the $Na^+$-$K^+$-$Cl^-$ transporter, NKCC1). The rise in the extracellular $K^+$ concentration following neuronal activation causes an astrocyte membrane depolarization, which is sensed by NKCC1 and causes an increase in the influx of both $Na^+$ and $HCO_3^-$. Their results further suggested that the activity of the $Na^+/K^+$-ATPase limited the ECS shrinkage by keeping the intracellular $Na^+$ concentration low, notably in the presence of an activity-induced increase of $Na^+$ influx. The low intracellular $Na^+$ concentration prevents intracellular osmolarity from reaching high levels, which in turn limits water influx from the ECS into the astrocyte.

In summary, the Østby et al. model describes ionic fluxes in the astrocyte and how they can impact the ECS volume during glutamatergic transmission. In several diseases, such as cortical spreading depression and epilepsy, the ECS and astrocyte volume is altered *(68)*. This model is best suited to study the impact of ionic fluxes in astrocytes on water and ionic homeostasis at synapses.

## 4.2. Model of potassium and sodium microdomains in astrocytes

Experimental studies showed the presence of ionic microdomains in thin astrocyte processes *(67)*, which correspond to small portions of the plasma membrane with inhomogeneous distributions of $Na^+$ channels and cotransporters, forming clusters. This spatial organization could result from spatially-restricted areas with negatively charged membrane lipids *(69)*. Breslin et al. *(64)* hypothesized that this localized negative charge might result in a slow diffusion of cations (positively charged ions) along the astrocyte processes. These localized negative charges create potential wells, characterized by a local minimum of potential energy. Potential wells restrict cation conduction, attracting and trapping the positive charges in the wells created by the negative charge.

To investigate their hypothesis, the authors proposed a multi-compartmental model of a synapse enwrapped by an astrocyte to explore the ion homeostasis in thin astrocyte processes and the interplay between the astrocyte and neuronal compartments (Figure 3). With the present model, the authors showed that the cation flow restriction forms a $K^+$ microdomain at the PAP, referred to as the perisynaptic cradle (PsC) in this study. Moreover, they showed that $K^+$ microdomains decrease the electrochemical gradient of $K^+$ and reduce the influx of $K^+$ through inward-rectifier $K^+$ channels ($K_{ir}$), facilitating the return to basal concentrations of $K^+$ in the perisynaptic ECS. Similar microdomains were observed for $Na^+$.

To further investigate the effect of such microdomains, Wade et al. *(67)* extended the Breslin et al. model by adding $Ca^{2+}$ channels onto the plasma membrane (Figure 3). Note that intracellular sources of $Ca^{2+}$ were omitted. The authors tested the hypothesis that $Ca^{2+}$ microdomains can be formed in the PsC and that those depend on $Na^+$ microdomains. $Na^+$ microdomains reversed the NCX, instigating an influx of $Ca^{2+}$ into the astrocyte. $Ca^{2+}$ microdomains in this case were not formed by potential wells but by the reverse mode of the NCX. Since $Na^+$ influx through EAAT-2 channels depends on presynaptic glutamate release, the Wade et al. model allows studying the effects of sustained neuronal activity on the intra- and extracellular ionic concentrations. The formation of $Na^+$ and $Ca^{2+}$ microdomains was itself sufficient to produce $Ca^{2+}$ transients, even in the absence of intracellular $Ca^{2+}$ stores.

In summary, Breslin et al. and Wade et al. were the first to hypothesize and simulate the formation of $Na^+$ and $K^+$ microdomains in PAPs and to test their effect on $Ca^{2+}$ microdomain activity. These models could help to study the effect of the modulation of the volume of astrocyte subcompartments on ionic microdomain formation and local ionic fluxes.

## 4.3. Model of $Ca^{2+}$ dynamics mediated by two different spatially-segregated pathways

While many computational models of astrocytes focus on one $Ca^{2+}$ pathway, Oschmann et al. *(65)* examined the interactions between i) glutamate-induced $Ca^{2+}$ signals and ii) glutamate uptake. In this model, it is assumed that the activity of glutamate transporters indirectly activates intracellular $Ca^{2+}$ influx through the activity of NCX, while $Na^{2+}$ and $K^+$ concentration gradients across the plasma membrane are maintained by the activity of the $Na^+/K^+$-ATPase. The model consists of a system of ODEs describing a single compartment of either an astrocyte process or soma divided into three compartments: a cylindrical ICS, a cylindrical ER (internal $Ca^{2+}$ store) within the ICS cylinder, and a cylindrical ECS, which has the same volume as the ICS (Figure 3). The model describes the dynamics of the astrocyte membrane potential, intracellular and extracellular ion concentrations ($Ca^{2+}$, $Na^+$, and $K^+$), and intracellular $IP_3$ concentration. Diffusion is not described as the model is not spatialized (see Section 6).

On one hand, this model assumes, based on previous experimental studies *(70, 71)*, that somatic $Ca^{2+}$ signals mostly result from mGluR activity-dependent $Ca^{2+}$ influx. The soma is characterized by a low surface-volume ratio and a high ER-ICS volume ratio ($ratio_{ER}$). On the other hand, $Ca^{2+}$ signals resulting from the activity of glutamate transporters are assumed to mostly occur near synapses, in PAPs, whose surface-volume ratio is high and $ratio_{ER}$ low. The authors used this model to investigate whether the activity of NCX and glutamate transporters can trigger $Ca^{2+}$ signals in PAPs. Intracellular $Ca^{2+}$ concentration in the PAPs only increased when the parameter value for the maximal pump rate of the NCX was increased. Blocking glutamate uptake by the astrocyte prevented $Ca^{2+}$ influx through the NCX.

Ziemens et al. *(72)* used the equation describing the NCX current from the Oschmann et al. model to predict that the increased $Na^+$ activity in PAPs measured experimentally upon NMDA application triggers NCX-dependent $Ca^{2+}$ influx (reverse mode) in PAPs.

In summary, the novelty of the Oschmann et al. model lies in the spatial separation within the astrocyte of mGluR- and glutamate transporter-dependent $Ca^{2+}$ signaling pathways. Therefore, the model is best suited for computational studies investigating the distinct $Ca^{2+}$ activity in the soma and PAPs.

## 4.4. Discussion

The maintenance of ion homeostasis is critical to ensure the propagation of action potentials in neurons and to prevent excitotoxicity. Several computational models have been developed to study ion homeostasis at tripartite glutamatergic synapses. For example, the model from Østby et al. can be used to study the interplay between ECS shrinkage, ion uptake, and water transport *(63)*. The Breslin et al. *(64)* and Oschmann et al. models *(65)* allow for studying the involvement of astrocytes in ion homeostasis and glutamate uptake. The Oschmann et al. model describes the interplay between two different $Ca^{2+}$ signaling pathways, while the Breslin et al. model allows for studying ionic microdomains in astrocyte leaflets and their effect on synaptic homeostasis. Altogether, these models can provide novel insights into the mechanisms by which astrocytes contribute to the regulation of ion homeostasis in the brain.

## 5. Metabolism

The idea that metabolic interactions occur between astrocytes and neurons has existed for more than a century now *(3)*. It is based on the observation that astrocyte processes are intimately juxtaposed to brain capillaries as well as neuronal synapses *(73)*. Astrocytes are involved in the uptake of glucose – the main energy source of the brain – from the blood and distribute it to other brain cells *(73)*. However, the detailed involvement of astrocytes in the metabolic processing of glucose remain unclear and controversial *(74)*. The major impediment in achieving a clear understanding of this phenomenon has been the subcellular resolution required to monitor metabolic fluxes during neuronal activity in the brain. This has been partially overcome by using more accessible experimental model systems like the retina, co-cultures of neurons and glial cells, as well as mathematical modeling *(75)*. Despite technological advances, the exact role of astrocytes and lactate in brain energy metabolism is still unresolved. Lactate is an alternative energy source to glucose. In the brain, lactate is produced by both astrocytes and neurons, which convert glucose into lactate through a process called aerobic glycolysis. Most investigations agree that lactate is transferred between astrocytes and neurons but disagree on the direction of this transfer: 1) astrocyte-to-neuron (ANLS) *(76)* or 2) neuron-to-astrocytes lactate shuttle (NALS) *(77)*.

Most of the existing computational models of astrocyte metabolism are based on the biophysical models proposed either by Aubert et al. *(78)* or Simpson et al. *(79)*. Both papers model the transport and processing of metabolites as a series of coupled differential equations that aim to explain experimental data. The second-generation models based on either of these models include recent insights and refined parameters (Figure 4) *(77, 80)*. Both models predict accurate glucose and lactate transients. However, despite a similar framework, they predict the opposite outcomes in the direction of lactate transfer. The differences seem to arise primarily from the way the models describe fluxes that are associated with i) the uptake of glucose by the astrocyte compartment from the basal lamina, ii) the uptake of glucose by the astrocyte compartment from the interstitium, and iii) the uptake of glucose by the neuronal compartment.

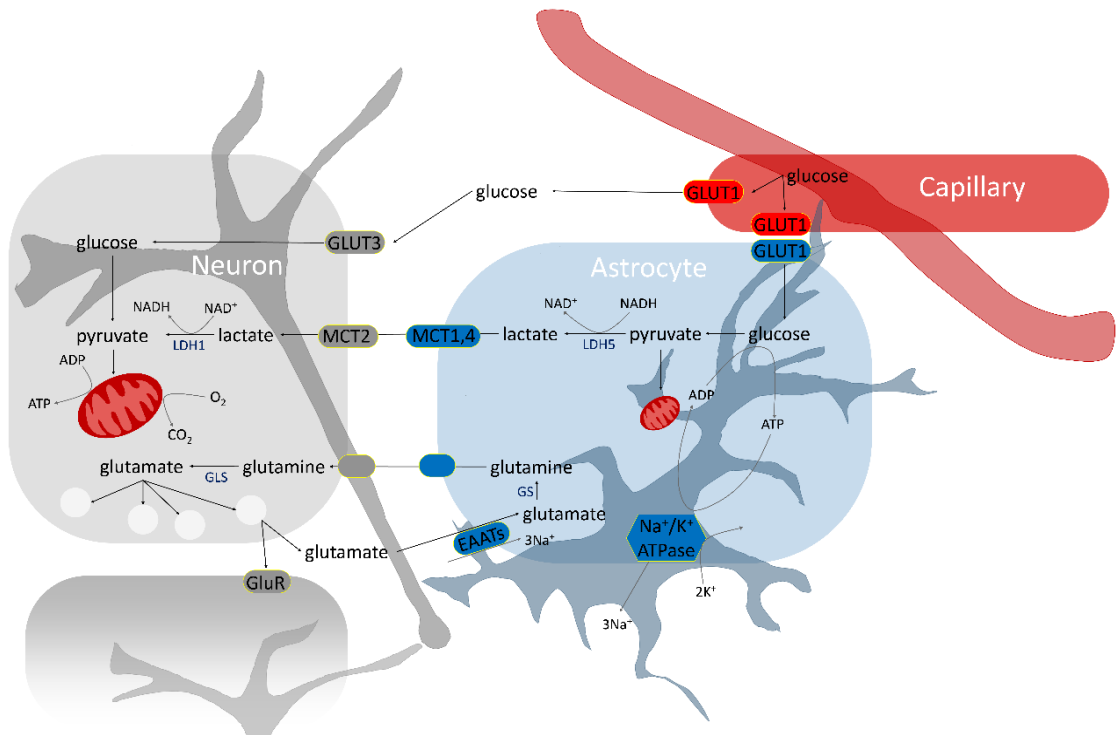

**Figure 4: Pathways described by models of astrocyte-neuron metabolic interactions.** Released glutamate from the presynaptic bouton can activate glutamate receptors (GluRs) at the astrocyte and postsynaptic membranes, which is an energetically expensive process. Following the uptake of glutamate by astrocytes, glutamate is converted into glutamine and transported to neurons, where it is converted back into glutamate by glutaminases (GLSs). In neurons, lactate can be used as an energy substrate following its conversion to pyruvate by the lactate dehydrogenase-1 (LDH1 ). Astrocytes and neurons take up glucose via GLUT1 (glucose transporter 1) and GLUT3 (glucose transporter 3), respectively.

## 5.1. Top-down model of the compartmentalization of metabolic pathways at tripartite synapses

Previous models that have attempted to resolve the ANLS versus NALS debate have relied on a bottom-up approach, wherein the energy needs of each of the biophysical processes involved in signaling were accounted for to calculate the total energy needs *(79)*. In contrast, the approach by Jolivet et al. *(76)* is a top-down approach that focuses on the energy that is available rather than required to understand the compartmentalization of different biochemical reactions involved in metabolic activity in neurons versus astrocytes. To that end, they used published datasets that describe the average tissue glucose and oxygen utilization at resting and active brain states *(81)*. They measured the linear relationship between the total cycling of neurotransmitters and the neuronal oxidative glucose utilization *(82)*. This allowed them to calculate the average tissue ATP production at rest and in the active state. Jolivet et al. then described the compartmentalization of neuronal and astrocyte oxygen and glucose metabolism (Figure 4) and used this information to investigate whether glucose is completely oxidized by these cells (based on the calculations by Gjedde et al. *(81)*). This method allows for a quantification of the energy budget of the brain constrained by *in vitro* experimental data and does not have to make any significant assumptions on parameter values. Their results suggest that a larger majority of glucose is taken up by astrocytes, while oxygen is mostly consumed by the neurons, and this consumption is correlated with neuronal activity. Although the model did not include glycogen, it was able to predict a wide range of *in vivo* data from the human brain. Their key finding that addresses the ANLS/NALS controversy is that astrocytes only oxidize a small portion of the glucose while neurons oxidize glucose-derived metabolites, which strongly supports the ANLS hypothesis since this results in the transfer of glucose-derived metabolites from astrocytes to neurons. The amplitude of this transfer goes up with increased neuronal activity.

In summary, the model quantifies the partitioning of the distribution of energy utilization, notably oxygen and glucose, by neurons and astrocytes. It provides a mathematical description of the neurovascular coupling at different spatial and temporal scales, describing the metabolic activity of neurons and astrocytes as well as the BOLD signal. The model can be used to predict the temporal dynamics of the consumption of lactate, glucose, and oxygen by the brain tissue.

## 5.2. Model of lactate and glucose levels in neurons and astrocytes during visual stimulation

The study by Mangia et al. *(77)* describes a mathematical model of NALS based on the Simpson et al. model *(79)* to gain insights into the compartmentalization of the metabolic activity of different brain cells. The model was implemented using *in vivo* data from the human brain, notably magnetic resonance spectroscopy, which quantified temporal changes in metabolite concentration during neuronal activity. The model simulates brain glucose and lactate levels in astrocytes and neurons (Figure 4), based on concentrations and kinetic rates measured experimentally. Parameters that govern the utilization of glucose and lactate by astrocytes and neurons were investigated over a wide range of values. Their results suggest that physiological parameter values predict NALS. Mangia et al. further demonstrate that ANLS is only possible under unrealistic conditions, where astrocytes display a twelve-times increased capacity for glucose transport and neurons do not respond to activation with increased glycolysis.

In summary, the main difference between Jolivet et al. *(76)* and Mangia et al. *(77)* as well as the rest of the ANLS and NALS models, seems to stem from the parameter values used to describe the amount of glucose that is entering the astrocyte compared to neurons. The NALS models keep the proportion of astrocyte glucose transport at around 20% *(77)* whereas the value of this parameter is more than 50% in ANLS models *(83, 84)*. Predictions from the Jolivet et al. and Mangia et al. models may need to be further tested in models of metabolic disorders or ischemic stroke to resolve the debate.

## 5.3. Discussion

Both ANLS and NALS models agree that glucose is partially transported into astrocytes from the blood *(85)*. The debate is about the proportion of this astrocyte glucose transport. The supporters of the ANLS models promote the idea that there is a shift in glucose utilization from neurons to astrocytes during glutamatergic activity. However, the modeling studies that support the NALS mechanism *(77, 86)* suggest that glucose and not lactate is the main energy substrate in the brain. This conclusion is based on the theoretical prediction that glucose transport capacity is larger in neurons than in astrocytes. The supporters of ANLS argue that this thesis is not consistent with the absence of a pathological phenotype in transgenic mice with decreased expression of the neuron-specific glucose transporter GLUT3 *(87)*. On the contrary, a decrease in the expression of GLUT1, the astrocyte-specific glucose transporter, leads to pathological conditions *(88)*. The strongest argument against the NALS model is the observation that the glucose utilization rate in neurons does not seem to vary depending on activity levels *(89)*. The NALS versus ANLS model debate thus remains unresolved to date.

# 6. Structure-function coupling

Astrocytes display a very complex nanoscopic morphology. Around 75 % of the total astrocyte volume consists of a meshwork of fine processes that are below the diffraction limit, thus unresolved by diffraction-limited light microscopy *(90)*. Such a complex cellular nano-architecture has been shown to greatly impact the function of various cell types. For example, the shape of dendritic spines controls the local sequestration of signals and thus strongly shapes synaptic function *(91–93)*. As modifying cell morphology without altering cell physiology is unfeasible experimentally, mathematical and computational modeling approaches are essential to investigate geometrical effects on cell signaling. Consequently, an increasing amount of astrocyte models describe and account for cell morphology and spatial effects. In this section, we present three different computational approaches and three models that can be used to study the effect of cell shape on astrocyte physiology at the single-cell level (Figure 5). Please refer to Section 7 for models taking into account the topology of astrocyte networks.

Before describing the three models, we provide a brief overview of modeling approaches taking into account the effects of cell geometry, referred to as spatially-extended approaches:
- Deterministic spatially-extended approaches describe the average behavior of populations of molecules within compartments. They are often referred to as compartmental models. Reactions within each compartment are described by ordinary differential equations.
- Stochastic spatially-extended approaches consider that molecular interactions are probabilistic events. There are two main stochastic approaches, described below. For more details, see *(94)*.
    1. Particle-based approaches, also referred to as particle-tracking or microscopic models, describe the position and state of all the molecules being modeled. Diffusing molecules are then tracked individually during simulation time.
    2. Voxel-based approaches, also referred to as population-based or mesoscopic, divide the modeled space into small compartments: voxels, most often cubes or tetrahedra. Each compartment is considered well-mixed and diffusion events are described as modifications of the number of molecules in two adjacent compartments.
- Hybrid approaches divide the system of interest into subcompartments, each describe by a different spatially-extended approach.

For more details on spatially-extended modeling techniques and tools, please refer to *(94–97)*.

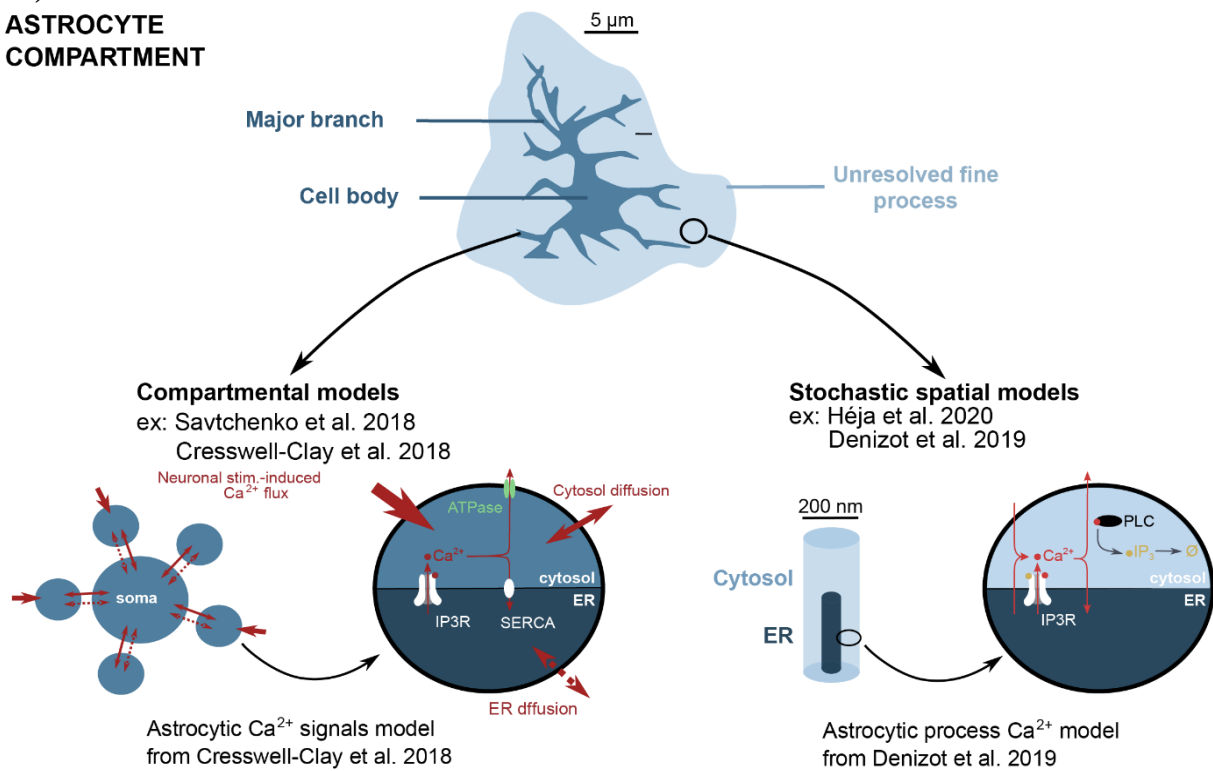

**Figure 5. Modeling strategies to investigate the effects of geometry on astrocyte activity depending on the compartment under study.** This schematic illustrates the main cellular subcompartments that characterize astrocyte morphology as well as examples of modeling techniques used depending on the compartment being modeled. Compartmental models such as implemented by Cresswell-Clay et al. *(98)* and Savtchenko et al. *(99)* are best suited to model $Ca^{2+}$ signal propagation and coupling within a whole astrocyte, while spatial models of fine processes such as developed by Héja et al. *(61)* and Denizot et al. *(100)* make it possible to study the effect of spatial factors such as cell morphology and $Ca^{2+}$ channels distribution on local microdomain activity.

## 6.1. ASTRO: a tool to simulate astrocyte activity in realistic astrocyte ultrastructures at the whole-cell level

ASTRO *(99)* is a computational tool for simulating astrocyte activity and is based on the NEURON software framework *(101)*. In addition to developing ASTRO, Savtchenko et al. *(99)* have developed an algorithm that creates geometries of single astrocytes based on experimental 3D reconstructions. The resulting geometries are adapted to be compatible with NEURON software, thus consisting of trees of 1D compartments. By adding processes randomly to those geometries, the ASTRO software allows the user to vary the tissue volume fraction occupied by processes as well as their surface-to-volume ratios. Reactions in each compartment are described by ODEs (deterministic spatially-extended approach, see above) and compartments are coupled by diffusion (Figure 5). The software has been extensively validated against experimental data performed on hippocampal astrocytes from the CA1 region, obtained with various approaches such as patch-clamp recordings, electrophysiology, two-photon excitation imaging, spot-uncaging, fluorescence recovery after photobleaching, *in vivo* $Ca^{2+}$ imaging, and quantitative correlational electron microscopy. Computational models of astrocytes have similar characteristics to those studied experimentally, such as their intracellular diffusional connectivity and their passive electrical properties.

ASTRO extends NEURON by adding features that are relevant for modeling astrocytes, such as the description of surface-volume ratios, sites of glutamate application, as well as the number and location of endfeet and gap junctions. Simulations using ASTRO have provided new insights into astrocyte activity, predicting various mechanisms controlling astrocyte physiology, such as the decrease of $Ca^{2+}$ wave speed and amplitude caused by increased $Ca^{2+}$ buffering. They further predicted that local $K^+$ efflux can efficiently prevent the spatial spread of elevations of intracellular $K^+$ concentration resulting from $K^+$ uptake. Finally, the detailed compartmentalization of the model allows for changing local characteristics of astrocyte activity, such as the local $Ca^{2+}$ channel cluster size. This allowed Savtchenko et al. to illustrate the complex interplay between the inter-$Ca^{2+}$ channel cluster distance, the associated $Ca^{2+}$ activity, and its fluorescence readout, mediated by $Ca^{2+}$ indicators.

The ASTRO tool can be used to test the effect of diverse characteristics of subcellular astrocyte subcompartments on cellular dynamics at the whole-cell level, such as membrane voltage spread, input resistance, and the generation of $Ca^{2+}$ waves.

## 6.2. A multicompartmental model of $Ca^{2+}$ activity in an astrocyte

Cresswell-Clay et al. *(98)* have developed a model that divides the astrocyte into different major compartments. Reactions involved in $Ca^{2+}$ signaling differ depending on the location within the cell and $Ca^{2+}$ diffuses between compartments. $Ca^{2+}$ can enter PAPs following neuronal stimulation. This reaction depends on $Ca^{2+}$ influx through $Ca^{2+}$ channels at the plasma membrane such as the NCX. Larger processes contain some ER and are characterized by $IP_3R$-dependent $Ca^{2+}$ signaling, including $Ca^{2+}$-induced-$Ca^{2+}$ release. $Ca^{2+}$ removal results from the activity of ATPases at the plasma membrane and the membrane of the ER (SERCA pumps). Larger compartments, i.e., the soma and five major branches, are non-spatial (single point models), connected to the rest of the astrocyte subcompartments by $Ca^{2+}$ diffusion in the cytosol and the ER (see Figure 5).

Cresswell-Clay et al. have used this model to study the influence of neuronal input properties, such as its amplitude or frequency, and diffusive properties, such as $Ca^{2+}$ diffusion coefficient in the cytosol or ER, on $Ca^{2+}$ spikes in the soma of the astrocyte. They found that concentrating neuronal inputs onto fewer astrocyte processes and increased synchrony of $Ca^{2+}$ signals in processes facilitated the emergence of somatic $Ca^{2+}$ spikes. Their results further suggested that cell morphology influenced $Ca^{2+}$ activity. In particular, an increased somatic volume was associated with a decreased somatic spike probability. Further, they found that an increased $Ca^{2+}$ diffusion coefficient in the cytosol facilitated the emergence of somatic spikes so that fewer process spikes were needed to result in a somatic event. Finally, $Ca^{2+}$ diffusion within the ER led to a non-monotonic variation of $Ca^{2+}$ somatic spikes with the neuronal input intensity, caused by Ca$^{2+}$ depletion in the ER for high neuronal input frequencies.

In summary, the Cresswell-Clay et al. model has improved our understanding of the integration of neuronal inputs by single astrocytes by varying spatial factors, such as the distribution of neuronal inputs over the astrocyte and diffusional properties in the cytosol and the ER, which cannot be performed experimentally. This model is best suited for studying the interactions between astrocyte compartments of different sizes in response to neuronal activity, notably the integration and propagation of Ca$^{2+}$ signals at the whole-cell level.

## 6.3. A spatial model of Ca$^{2+}$ activity in a perisynaptic astrocyte process

Most astrocyte-neuron communication occurs in fine PAPs that contain a very low number of molecules and ions so that the kinetics of the associated reactions are highly stochastic. Stochastic spatially-extended approaches are best suited to model astrocyte physiology at this spatial scale. The model from Denizot et al. *(100)* corresponds to a model of IP$_3$R-dependent Ca$^{2+}$ signals in fine processes (Figure 5), implemented both in 2D, with a custom-made particle-based simulator, and in 3D, using the voxel-based STEPS software *(102)*. The latter allows running simulations at the nanoscale in both simplified 3D shapes of thin processes and more realistic ultrastructures reconstructed from electron microscopy, for example. Simulations of a non-spatial, non-stochastic implementation of the model highlighted that stochasticity was necessary for spontaneous Ca$^{2+}$ signals to be triggered in fine processes. The 2D implementation allowed the authors to explore the range of dynamical behaviors that the model displays, suggesting that Ca$^{2+}$ peak frequency increases when Ca$^{2+}$ channels are organized into spatial clusters. Simulations of the 3D model implementation were performed in a simplified astrocyte process morphology in 3D, consisting of a 1 μm-long, 100 nm in radius cylinder, which displayed a similar Ca$^{2+}$ microdomain activity than recorded in organotypic cultures of hippocampal astrocytes. Simulations quantified the alteration of Ca$^{2+}$ signals by Ca$^{2+}$ indicators, which are necessary to perform Ca$^{2+}$ recordings experimentally. Increased concentrations of Ca$^{2+}$ indicators were notably resulting in a decrease in Ca$^{2+}$ peak amplitude and frequency.

The model from Denizot et al. has recently been used to investigate the effect of remodeling the astrocyte nano-architecture observed in pathological hypo-osmotic conditions *(103)* on local astrocyte Ca$^{2+}$ activity at tripartite synapses *(104)*. Simulation results suggest that the nanoscale reticular morphology of astrocyte processes observed in healthy tissue *(105)* enhances local Ca$^{2+}$ activity and that this effect is hindered in pathological conditions, which was confirmed by Ca$^{2+}$ imaging experiments. More recently, simulations of this model in realistic 3D geometries of PAPs reconstructed from electron microscopy gave new insights into the complex interplay between ER shape and distribution, the clustering of Ca$^{2+}$ channels, and Ca$^{2+}$ buffering mechanisms in regulating microdomain Ca$^{2+}$ activity at tripartite synapses *(106)*.

The high spatial resolution of this model comes at a high computational cost and simulations of hundred seconds of chemical reactions in a fine process take several days to compute, which is much slower than the compartmental models of Savtchenko et al. *(99)* and Cresswell-Clay et al. *(98)*, despite simulating smaller subcellular compartments. For that reason, the model from Denizot et al. is best suited to study astrocyte physiology in fine processes and can be used to test the effect of spatial factors, such as cell morphology and the distribution of $Ca^{2+}$ channels, on astrocyte microdomain $Ca^{2+}$ activity.

## 6.4. Discussion

In this section, we have presented some of the recent models that take into account the complex morphology of astrocytes to investigate the effect of spatial properties of astrocytes on their activity. The presented models describe different signaling pathways and cell shapes, using different spatial resolutions and accuracy. Such spatially-extended models are useful tools to test the effect of factors that might be crucial to astrocyte physiology, such as the location and density of gap junctions, the distribution and size of $Ca^{2+}$ channel clusters, and the local variability of astrocyte morphology. As those parameters vary drastically in pathological conditions and are often inaccessible experimentally, those models offer valuable opportunities to better understand the biochemical processes that underlie astrocyte activity and astrocyte-neuron communication in health and disease.

# 7. Astrocyte networks

Astrocytes establish complex networks with the numerous cells they are contacting. Notably, in the human brain, a single protoplasmic astrocyte could contact up to 2 million synapses residing in its territorial domain *(107)*, forming numerous tripartite synapses (see Section 2) *(108)*. Astrocyte and neuronal networks are thus tightly interwoven. Astrocyte activity is correlated to neuronal synchronization *(109, 110)*. Yet, the mechanisms by which neuron-astrocyte communication shapes network activity in health and disease remain poorly understood.

Astrocytes are characterized by non-overlapping spatial domains *(111)* and are connected to neighboring astrocytes through gap junction channels. This coupling allows for the flow of small molecules (e.g., $IP_3$) and ions from one cell to another *(112)*. Intercellular $Ca^{2+}$ waves can spread from one astrocyte to up to 70 neighboring cells in rodent cultures *(113)*. Such a spatial spread of astrocyte $Ca^{2+}$ signals can influence the activity of numerous neuronal circuits simultaneously. Studying the interplay between astrocyte and neuronal activity at the network level is thus crucial to expanding our understanding of brain physiology.

In this section, we present three computational network models that describe astrocyte-neuron networks that take into account the network topology and include hundreds of astrocytes (Figure 6).

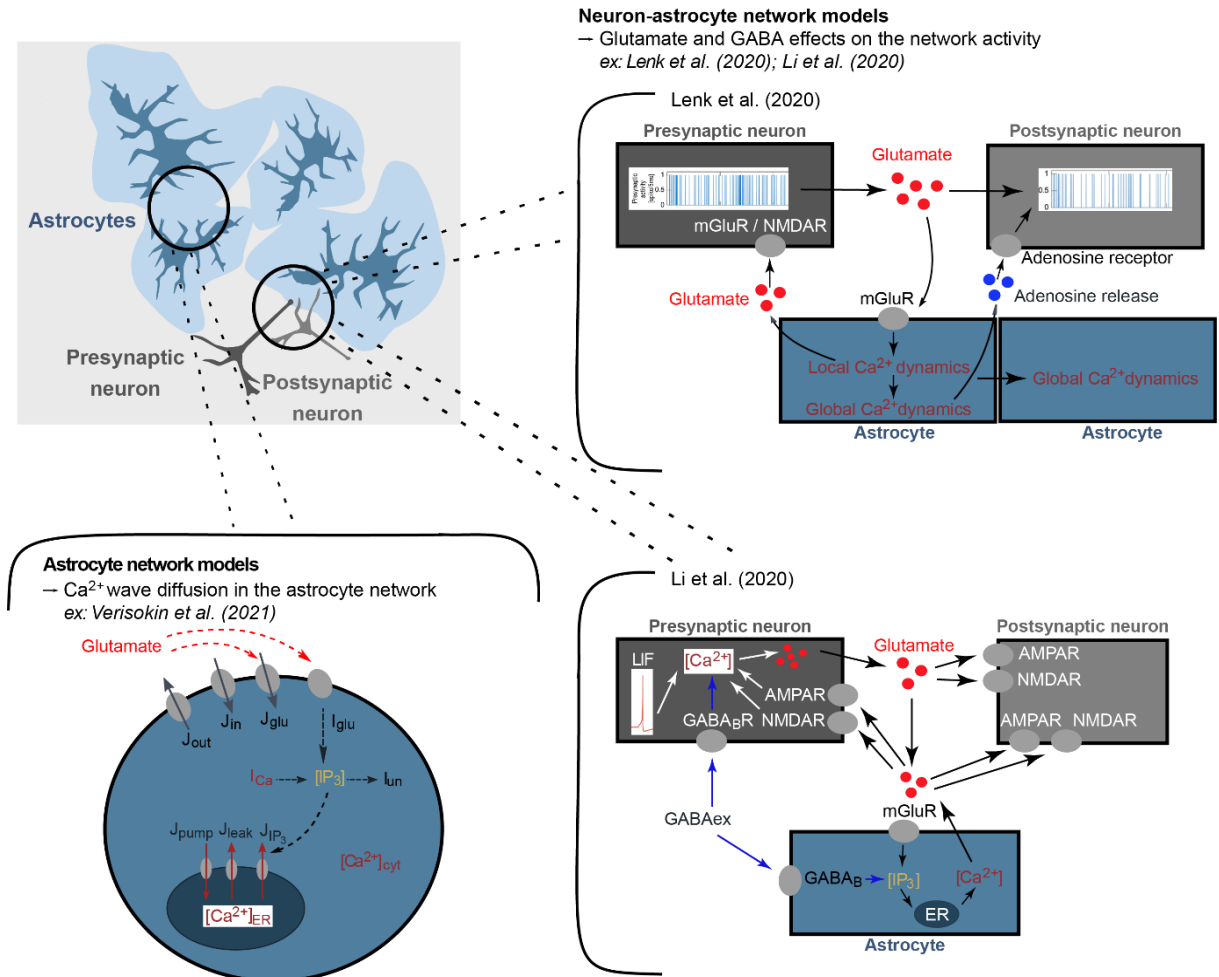

**Figure 6: Astrocyte and neuron-astrocyte network models.** Verisokin et al. *(114)* modeled the propagation of Ca$^{2+}$ waves in astrocyte networks. Both Lenk et al. *(34)* and Li et al. *(115)* models describe the propagation of Ca$^{2+}$ waves in astrocyte networks coupled with the excitatory and inhibitory transmission in neighboring neuronal networks.

### 7.1. A topologically realistic model of astrocyte networks

Verisokin et al. *(114)* have implemented a morphologically-detailed network model of astrocytes focusing on the spatial spread of Ca$^{2+}$ signals (Figure 6). Each astrocyte shape was generated by randomly transforming confocal microscope images. In the model, astrocytes were randomly placed onto a grid of 0.275 μm/px for the single astrocyte simulations and 0.55 μm/px for the network simulations, while ensuring that astrocyte territories did not overlap. In the model, each astrocyte is subdivided into i) the soma with thick branches, ii) the thin astrocyte processes, and iii) the surrounding ECS. In the model, the astrocytes are stimulated by glutamate released by connected presynaptic neurons, modeled with a stochastic Poissonian distribution. In other words, the neurons are not explicitly modeled. Astrocyte activity was described using the mGluR-dependent Ca$^{2+}$ signaling model by Ullah et al. *(116)*. The contribution of the different Ca$^{2+}$ pathways (i.e., ER- or plasma membrane-mediated) varied depending on the surface-volume ratio of the subcellular compartment. More precisely, ER-mediated Ca$^{2+}$ signals prevail in the soma and thick branches while signals in leaflets are mediated by channels at the plasma membrane. Ca$^{2+}$ and IP$_3$ diffusion are described both within and between astrocytes.

The model was able to reproduce Ca$^{2+}$ activity measured experimentally in terms of duration and spatial spread at the single-cell level as well as in terms of spatial spread in the network *(117)*. Simulation results indicated that, even though all cells were described similarly, the network presented a pacemaker-like behavior, i.e., the spread of signals originated from a specific pool of cells. This resulted from differences in cell morphology as well as in the astrocyte-to-astrocyte contacts in the network and drove the activation of multicellular Ca$^{2+}$ waves, which often displayed similar spatiotemporal properties.

The novelty of the Verisokin et al. model is that it simulates the astrocyte Ca$^{2+}$ activity within realistic cell shapes. This feature makes the model most suitable to study the effect of cell morphology on Ca$^{2+}$ dynamics at the whole cell as well as at the astrocyte network levels.

### 7.2. A topologically realistic model of neuron-astrocyte networks

Lenk et al. *(34)* introduced a neuron-astrocyte network model with a biologically-plausible network topology. The simulations aimed to reproduce neuronal spiking recorded from rodent co-cultures plated on *in vitro* microelectrode arrays. In the model, the network includes 250 neurons and varying ratios of astrocytes, which are distributed over a 750 × 750 μm$^2$ 2D space. Both cell types are modeled as points in space, i.e., they do not have a morphology. The astrocytes are randomly placed in the 2D space and are connected via gap junctions if the inter-soma distance between two cells is lower than 100 μm. The neurons, thereof 80% excitatory and 20% inhibitory, are randomly distributed in the 2D space, and long-distance connections (up to 500 μm) are allowed.

Of note is that the model currently only connects excitatory neurons with astrocytes due to the limited information on interactions between inhibitory neurons and astrocytes at the time. Upon incoming spikes, the excitatory neurons release glutamate into the synaptic cleft, which activates glutamate receptors at the membrane of the postsynaptic neuron and the perisynaptic astrocyte. The astrocyte activity is modeled by Lenk et al. using the model from De Pittà et al. *(27)* (see Section 2.1), to which they added the release of the gliotransmitters glutamate and ATP by the astrocyte into the synaptic cleft (Figure 6).

In the first set of *in silico* experiments, each excitatory presynaptic neuron was connected to an astrocyte, while astrocytes were not coupled. This network topology resulted in increased spike and burst rates to pathological levels. Then, simulations were performed in the complete neuron-astrocyte network model with 10, 20, or 30% astrocytes, which were connected by gap junctions. This topology led to a reduction of the neuronal spiking and bursting rates to healthy ranges. Increasing the number of astrocytes shaped neuronal network activity by preventing overexcitation.

In Fritschi et al. *(118)*, the Lenk et al. model was used to investigate four hypotheses on the pathological mechanisms involving astrocytes in schizophrenia: i) The number of neurons or astrocytes in the network is reduced. ii) There is an effect of astrocyte ATP on postsynaptic activity. iii) The release of glutamate from the presynapse and the uptake of glutamate by the astrocyte is altered in schizophrenia. iv) The excitatory and/or inhibitory synaptic strength, i.e., the coupling between neurons is stronger in this disease.

In summary, Lenk et al. modeled the communication between neurons and astrocytes in networks with a high number of cells. The network topology is highly controllable by the model user, which makes the model useful to study the effect of network topology on neuronal and astrocyte activity in 2D and 3D *(119–121)*.

### 7.3. A network model of GABA-evoked neuron-astrocyte communication

Li et al. *(115)* developed a model of extracellular GABA (γ-aminobutyric acid) activation of astrocytes, resulting in IP$_3$-mediated Ca$^{2+}$ signals (Figure 6). The neurons are modeled using the leaky integrate-and-fire (LIF) formalism *(122)*, which describes a neuron as an electrical circuit composed of a capacitor (C) in parallel with a resistor (R), such as in an RC circuit. When a current is injected into the model, the LIF neuron acts as a resistor, and, in the absence of an input current, the membrane potential discharges exponentially to its resting value. Excitatory and inhibitory neurons differ in the model only by their initial excitatory and inhibitory conductances. The presynaptic neurons express glutamate receptors (e.g., NMDARs and mGluRs) and GABA receptors (e.g., GABA$_B$Rs). The model further describes the activity of NMDA and AMPA receptors in the postsynaptic neurons.

The astrocytes express mGluRs at their plasma membrane, whose activation leads to IP$_3$-evoked Ca$^{2+}$ signaling (see Section 2). The novelty of this model is to incorporate GABA$_B$Rs at the astrocyte plasma membrane, whose activation also results in IP$_3$ synthesis. The 2D network model comprises 500 neurons (400 excitatory and 100 inhibitory), with a 20 % connection probability, and 400 astrocytes. The cells are uniformly distributed onto a 10 x 10 mm$^2$ planar grid. The astrocytes in the network are on average connected to 100 neighboring excitatory synapses and four astrocytes. The model simulates the response of the network to an injection of exogenous GABA in the ECS. At the synaptic level, the model describes the activation of GABA$_B$Rs in the astrocyte and the presynaptic neuron. In the presynaptic neuron, the activation of GABA$_B$Rs decreases glutamate release probability, which counteracts the increased glutamate release from the astrocyte. Changes in GABA concentrations are based on the presynaptic release and exogenous input and then decay exponentially. The results of this work suggest how elevated extracellular GABA concentrations can increase the duration and amplitude of astrocyte Ca$^{2+}$ signals in a concentration-dependent manner. Without external GABAergic stimuli, the astrocyte Ca$^{2+}$ oscillations were slower and more similar to those measured in healthy conditions.

Overall, the Li et al. model describes the effects of GABA release on glutamatergic synaptic transmission and is thus suitable for studying the interplay between excitatory and inhibitory signaling in neuron-astrocyte networks.

### 7.4. Discussion

Astrocytes and neurons in the brain form interconnected networks. Verisokin et al. *(114)* modeled the influence of cell morphology on Ca$^{2+}$ activity in astrocyte networks. The model framework is similar to Savtchenko et al. *(99)* (Section 6.1) but with fewer biochemical details, thus facilitating the simulation of astrocyte activity at the network level. Li et al. *(115)* and Lenk et al. *(34)* models describe the interactions between neurons and astrocytes. These models do not describe cell morphology but rather concentrate on neuron-astrocyte communication through neuro- and gliotransmission at the network level. This section illustrates how computational network models can be used to test different hypotheses related to gliotransmission and its effect on neuronal activity (see *(22, 23)* for reviews on the current debates).

# 8. Concluding remarks

One of the biggest unresolved questions in neuroscience lies in understanding the physiological roles played by glial cells, the 'other half of the brain', in different anatomical regions and brain states. By now, it is well established that astrocytes, the most abundant glial cell type, display a rich repertoire of functions that operate over diverse spatiotemporal scales. Since most of the 'currency' of these cells (i.e., glutamate, ATP, $Ca^{2+}$ signals, etc.) is common to that of neurons, disambiguating their precise contribution to brain function in health and disease has been challenging. The history of neuroscience tells us that groundbreaking discoveries have often materialized through a synergy between experimental insights and mathematical and computational models *(123–127)*. In the last two decades, we have witnessed a deluge of experimental investigations targeting astrocytes, which has helped deepen our understanding of their contribution to brain function. However, making sense of the resulting high-dimensional data is a major challenge, so that a modeling framework equivalent to that of neurons is critical to fill in the missing gaps. A model is an abstraction with an immediate goal to reduce the dimensionality of the problem under consideration. Thus, this minimal representation of the system carries information about the components that are critically involved in a function of interest. The iterative and trial-and-error process of building minimal representations (or models) is typically based on a key observation from experiments and a modeling intuition (a good example is the Hodgkin-Huxley model of action potential propagation *(38)*). In this scheme, variables, timescales, and parameters can be systematically explored; what does not fit is thrown out and new components are brought in. Computational models of astrocytes range widely in scale; from the nanoscopic interactions of individual molecules to intercellular processes at the network level. Besides the spatial scale, the temporal scales of these models also vary largely (from milliseconds to seconds).

We envision the reader selecting sections and models that are relevant to the experiments they are running and the data obtained. We further aim to provide a concise overview of the types of models that are available, together with a glimpse window into their usage. Our aim with this book chapter is to highlight how computational models complement experiments in the quest for unraveling neuron-astrocyte communication at glutamatergic synapses to foster collaboration between neuroscience disciplines.

# 9. List of resources

ModelDB (https://senselab.med.yale.edu/ModelDB/) and CellML (https://www.cellml.org/) are the main open-access databases that host numerous models of neurons and astrocytes. Table 1 provides the links to the models described in this chapter that are available online.

**Table 1:** Websites with the code of available models

| Section | Model | Website |
|---|---|---|
| 2.1. | De Pittà et al. | https://github.com/mdepitta/comp-glia-book/tree/master/Ch5.DePitta |
| 3.1. | Rădulescu et al. | https://github.com/scimemia/Glutamate-transporters-estimates |
| 4.1. | Østby et al. | https://models.physiomeproject.org/exposure/d9de93b128da322a4d50f24589980ea1/ostby_oyehaug_einevoll_nagelhus_plahte_zeuthen_voipio_lloyd_ottersen_omholt_2008.cellml/view |
| 4.3. | Oschmann et al. | (partially available) https://github.com/FranziOschi/AstroMultiComp |
| 6.1. | Savtchenko et al. | https://senselab.med.yale.edu/ModelDB/ShowModel?model=243508#tabs-1 |
| 6.2. | Cresswell-Clay et al. | https://github.com/FSUcilab/Compartmental_model_astrocytes |
| 6.3. | Denizot et al. | https://senselab.med.yale.edu/ModelDB/ShowModel?model=247694#tabs-1 |
| 7.1. | Verisokin et al. | https://zenodo.org/record/4552726#.Yrr28C8RppR |
| 7.2. | Lenk et al. | https://github.com/kerstinlenk/INEXA_FrontCompNeurosci2020 |

# Acknowledgement

K.L.'s research was partially conducted while visiting the Okinawa Institute of Science and Technology through the Theoretical Sciences Visiting Program. A. D.'s work was funded by the Okinawa Institute of Science and Technology Graduate University and by the Japan Society for the Promotion of Science Postdoctoral Fellowship for Research in Japan (Standard, P21733).